\documentclass[utf8]{frontiersFPHY} 

\setcitestyle{square} 
\usepackage{url,hyperref,lineno,microtype,subcaption}
\usepackage[onehalfspacing]{setspace}

\def\keyFont{\fontsize{8}{11}\helveticabold }
\def\firstAuthorLast{Love et al.} 
\def\Authors{Teri Love \,$^{1,*}$, Thomas Neukirch\,$^{1}$ and Clare E. Parnell\,$^{1}$}
\def\Address{$^{1}$School of Mathematics and Statistics, University of St Andrews, Scotland}
\def\corrAuthor{Teri Love}

\def\corrEmail{tl28@st-andrews.ac.uk}

\begin{document}
\onecolumn
\firstpage{1}

\title[Analysing Flare Observations using CNNs]{Analysing AIA Flare Observations using Convolutional Neural Networks} 

\author[\firstAuthorLast ]{\Authors} 
\address{} 
\correspondance{} 

\extraAuth{}

\maketitle

\begin{abstract}

\section{}
In order to efficiently analyse the vast amount of data generated by solar space missions and ground-based instruments, modern machine learning techniques such as decision trees, support vector machines (SVMs) and neural networks can be very useful. In this paper we present initial results from using a \textit{convolutional neural network} (CNN) to analyse observations from the \textit{Atmospheric Imaging Assembly} (AIA) in the 1600$\mathring{A}$ wavelength. The data is pre-processed to locate flaring regions where flare ribbons are visible in the observations. The CNN is created and trained to automatically analyse the shape and position of the flare ribbons, by identifying whether each image belongs into one of four classes: two-ribbon flare, compact/circular ribbon flare, limb flare or quiet Sun, with the final class acting as a control for any data included in the training or test sets where flaring regions are not present. The network created can classify flare ribbon observations into any of the four classes with a final accuracy of 94\%. Initial results show that most of the images are correctly classified with the compact flare class being the only class where accuracy drops below ~90\% and some observations are wrongly classified as belonging to the limb class.

\tiny
 \keyFont{ \section{Keywords:} convolutional neural network, solar flares, flare ribbons, machine learning, classification, helio19} 
\end{abstract}

\section{Introduction}
\label{sec:introduction}

The steady improvement of technology and instrumentation applied to solar observations has led to the generation of 
vast amounts of data, for example the \textit{Solar Dynamics Observatory} (SDO) collects approximately 1.5 terabytes of data everyday  \cite{2012SoPh..275....3P}. The analysis of these data products can be made much more efficient by the use of modern machine learning
techniques such as decision trees, support vector machines (SVMs) and neural networks. In this paper we describe some
initial results we obtain using a convolutional neural network to analyse SDO data.
Basic applications of CNNs to solar physics data classification is shown in \citep[e.g.][]{Armstrong2019, deepcnn_events}, however CNNs have also started being applied to the prediction of solar events, in particular flares and CMEs, that can affect space weather 
as considered, for example, by \citet{bobra2015}, \citet{cnn_goes}, \citet{sunspotcnn}.

In this paper we focus on solar flares and in particular on the classification of the morphology of flares displaying visible flare ribbons
\citep[e.g.][]{1989SSRv...51...49K,fletcher2001}.
Throughout this paper, flare observations from the \textit{Atmospheric Imaging Assembly} (AIA) \cite{AIA} onboard SDO were used, specifically AIA 1600 $\mathring{A}$. These observations clearly show the flare ribbons as they appear on the solar surface. 

The locations and shapes of flare ribbons are thought to be closely linked to the geometry and 
topology of the solar magnetic field in the flaring region. For example,
the 
ribbon shapes and lengths have been connected to the presence of separatrix surfaces and 
quasi-separatrix layers (QSLs)
\citep[e.g.][]{2000ApJ...540.1126A, 2015ApJ...810...96S, 2019ApJ...871....4H, 2016A&A...591A.141J}.
The ribbon shapes found and analysed throughout 
these studies are mostly two-ribbon flares with two 'J' shaped ribbons, however 
it is known that other ribbon shapes can also occur with circular or compact flare ribbons also being observed.
One motivation of the work presented in this paper is to create a tool that allows the classification of large data sets to generate a catalogue of flares associated with their ribbons, which could automatically be detected and classified. The catalogue could then, for example, be used in connection with magnetic field models to obtain
better statistics on the possible correlation of ribbon geometry and magnetic field structure.

This paper considers all C, M and X class flares 
\citep[see e.g.][for a definition of GOES classes]{Fletcher2011} that occurred 
between November 2012 and December 2014 and attempts to classify the shape of all observable flare ribbons. 
To do this a \textit{convolutional neural network} (CNN) consisting of two hidden layers was 
created and trained to predict four classes of ribbons and flares. These four classes are 
two-ribbon flares, limb flares and circular/compact ribbon flares, with the fourth class acting as 
a control class to process quiet Sun images that may also be processed through the CNN. 
The network was trained on a dataset containing 540 images (including validation images), and was tested using an unseen dataset containing 430 images.

The paper is structured as follows. In section \ref{sec:methods} we describe the design and training of our CNN.
The preparation of the data used in the paper is discussed in section \ref{sec:dataprep}, our results are presented
in section \ref{sec:results} and we conclude with a discussion of our findings in section \ref{sec:discussion}.

\section{Methods}
\label{sec:methods}
Convolutional neural networks (CNNs) are a type of machine learning technique commonly used to find patterns in data and classify them. Instead of being given explicit instructions or mathematical functions to work they use patterns and trends in the data, initially found through a ``training data" set. This data set should be the set of inputs for the CNN - usually a subset of the data that one would initially want to classify or detect. This allows the network to ``learn" the patterns and trends such that it can independently classify unknown data.

\subsection{CNN Design}
To create a basic CNN there must be at least 3 layers; an input layer, a hidden layer and an output layer  \citep[e.g. ][]{10.5555/109230.109279, 2012arXiv1207.0580H, imagenet_cnn, review_cnns, Szegedy_2015_CVPR}. The input layer is the first initial network layer which accesses the data arrays inputted into the model which are to be trained upon. The data input has usually been through some pre-processing before being used by the network, the pre-processing used on the AIA data is discussed in Section \ref{sec:dataprep}. 

The hidden layer is a convolutional layer where instead of applying a layer using matrix multiplication, as in general neural networks, a layer using a mathematical convolution is used instead. Although this is the basic set-up for a CNN, most CNNs have multiple hidden layers before having a fully connected output layer. The different types of hidden layers that can be used are: convolutional, pooling, dropout (\citet{2012arXiv1207.0580H}) and fully connected layers. The final output layers are usually built from fully connected (\textit{dense}) layers. These layers take the output from the hidden layers and process it such that for each data file a pre-defined class is predicted by the network. 


A convolutional layer performs an image convolution of the original image data using a kernel to produce a \textit{feature map}. These kernels can be any size but are commonly chosen to be of size $3\times3$. The stride of the kernel can also be set in the convolutional layers indicating how many pixels it should skip before applying the kernel to the input - this has been set as 1 for the CNN here such that the kernel has been applied to every pixel in the input. If larger features were to be classified larger strides could be used.

The kernel moves over every point in the input data, producing a single value for each 3x3 region by summing the result of a matrix multiplication. The value produced is then placed into a feature map which is passed onto the next layer. As the size of the feature map will be smaller than the input, the feature map is padded with zeros to ensure the resulting data is the same size as the original input.
After the feature map is produced the convolutional layer has an associated \textit{activation function} which produces non-linear outputs for each layer and determines what signal will be passed onto the next layer in the network. A common activation function used is the \textit{rectified linear unit} (ReLU \citep{relu}) , which is defined by;
\begin{equation}
    f(x) = max (0, x).
    \label{eq:relu}
\end{equation}
Other activation functions such as \textit{linear, exponential} or \textit{softmax} (see Equation \ref{eqn:softmax}) can also be implemented, however for the convolutional layers in our model only ReLU is used, as the function can be calculated and updated quickly with gradients remaining high (close to 1), with ReLU also avoiding the vanishing gradient problem.

Although convolutional layers make up the majority of the hidden layers within a CNN, other hidden layers are also important to avoid over-training of the network. Implemented after convolutional layers, \textit{pooling} layers are commonly used to deal with this. Pooling layers help to reduce over-fitting of data and reduce the number of parameters produced throughout training - which causes the CNN to learn faster. The most common type of pooling is \textit{max pooling} which takes the maximum value in each window and places them in a new and smaller feature map. The window size for the max pooling layers can be varied similarly to the convolutional kernels, however throughout this paper all max pooling layers had a kernel of size 2x2. Although the feature map size is being reduced, the max pooling layers will keep the most important data and pass it onto the next training steps. 

For the CNN created to analyse the flare ribbons, two convolutional layers were implemented after the input layer. These layers were both followed by max-pooling layers with a stride of 2. Both layers were implemented using ReLU activation functions, however the first convolutional layer had 32 nodes whereas the second layer was implemented with 64 nodes before being passed onto fully connected layers.

Once the convolutional and pooling layers have been implemented as hidden layers, the final feature map output is passed onto output layers which allows the data to be classified. These classification layers are made up of fully connected (FC) layers - similar to those in a normal neural network. FC layers only accept one-dimensional data and so the data must be flattened before being passed into them. The neurons in the FC layers have access to all activations in previous layers - this allows them to classify the inputs. The final fully connected layer should have the number of classes as its units, with each output node representing a class.

An additional output layer that can be implemented before a FC layer is a \textit{dropout layer}. This layer is implemented before a FC layer to indicate that random neurons should be ignored in the next layer i.e. they have \textit{dropped out} of the training algorithm for the current propagation. Hence if a FC layer is indicated to have 10 neurons, a random set of these will be ignored when training \citep[see e.g.][for further information on dropout layers]{2012arXiv1207.0580H}.

\begin{figure}[ht!]
\begin{center}
\includegraphics[width=12cm]{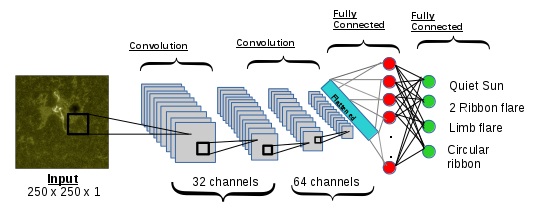}
\end{center}
\caption{ Layout of CNN created, including two convolutional and max pooling layers and two fully connected layers. The first convolutional layer has 32 channels followed by max pooling layer and the second convolutional layer has 64 channels followed by a max pooling layer. The fully connected layer has 128 nodes and then the final fully connected layer has four nodes which correspond to each of the classes - Quiet Sun, two-ribbon flares, limb flares and circular/compact flares.}\label{fig:cnn_layout}
\end{figure}

The CNN was created and trained using Keras \cite{chollet2015keras}, with the network layout shown in Figure \ref{fig:cnn_layout}. This shows the two convolutional and pooling layers previously discussed, with a dropout layer implemented before the data is passed onto two FC layers, with  128 and 4 nodes respectively. A breakdown of all parameters used in each layer are shown in Table \ref{table:parameters}.

\subsection{Model Training}
\label{training}
The previous section described the basic design of the CNN used throughout this paper. Here we will describe the training process carried out on the model. 

When data is passed through the network, at each layer a loss function is used to update the model weights. This loss function carries out the process known as \textit{back-propagation} \cite{backprop}, where differentiation takes place and the network learns the optimal classifications for each training image. The loss function chosen for our model is known as \textit{categorical cross entropy}. This cross entropy loss is calculated as follows;
\begin{equation}
    CEP = - \sum_{c=1}^{M}y(x_{i})log(p(x_{i})),
\end{equation}
where $M$ is the number of classes (here $M=4$) and y is the binary indicator (0 or 1) such that if $y=1$ the observation belongs to the class and $y=0$ if it does not. Finally $p$ is the probability that the observation belongs to a class, $c$.

The probability, $p(x_{i})$, of each class is calculated using a softmax distribution such that;
\begin{equation}
    p(x_{i}) = \frac{e^{x_{i}}}{\sum_{k}e^{x_{i}}}.
    \label{eqn:softmax}
\end{equation}
This function should tend towards 1 if an observation belongs to a single class and tends to 0 for the other 3 classes to indicate that the network does not recognise it as belonging to those classes. The resultant classification is selected by choosing the largest probability that lies above $p(x_{i}) = 0.5$.

The network is trained on 540 1600$\mathring{A}$ AIA images. The data processing is discussed in Section \ref{sec:dataprep}, with each image used containing a single flare, unless it belongs to the quiet Sun class. The four classes are as follows:
\begin{enumerate}

\item \textbf{Quiet Sun}

No brightenings present on the surface, hence should give an indication of general background values. (It should be noted that none of these observations are taken on the limb.)

\item \textbf{Two-ribbon Flare}

Two flare ribbons must be clearly defined in the observations. However the shape does not matter here e.g. if there are 2 semi circular ribbons the flare is classified as a two-ribbon flare and not a circular flare.

\item \textbf{Limb Flares}

The solar limb must be clearly observed in this snapshot observation with a flare brightening being visible. The limb class was chosen to start at a specific distance from the solar limb to reduce confusion with other classes. This will be discussed further in Section \ref{sec:results}.

\item \textbf{Circular Flare Ribbons}

Here a circular ribbon shape of any size must be observed. It should be ideally a singular ribbon so as not to be confused with the two-ribbon flare class. Compact flares were also included here, they appear in the data as round 'dot' like shapes. 
\end{enumerate}

Classes were divided almost evenly to stop observational bias from entering the model during training and although there is a slight class imbalance it is not large enough to affect the accuracy of the model. From the training set used, 40\% of the data was used as a validation data set with the remaining 60\% used to train the model. The learning rate chosen was $10^{-4}$ with a batch size of 32 selected for both training and validation to allow the use of mini-batch gradient descent throughout training. Although larger batch sizes would speed up the training process, to get better generalisation of the model a smaller batch size was picked to improve the model accuracy.

\begin{figure}[ht!]
\begin{center}
\includegraphics[scale=0.45]{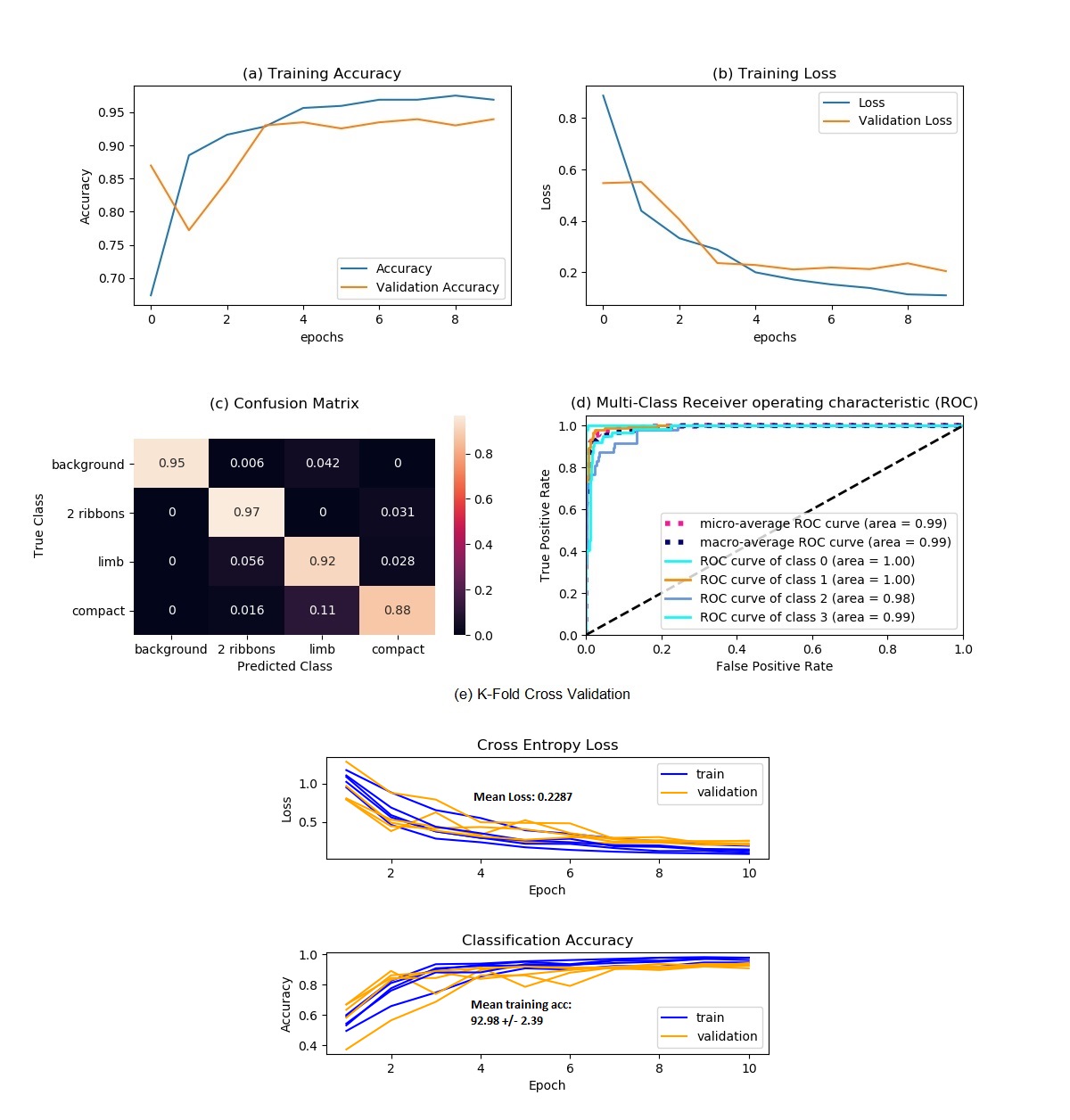}
\end{center}
\caption{(a) Training accuracies with both validation and training accuracies shown over 10 epochs; (b) Training and validation loss shown over 10 epochs; (c) Shows the confusion matrix created on the test set, with the diagonal showing the correctly identified ribbon types; (d) shows the receiver operating characteristic (ROC) curve which has been modified to include a curve for each class and the micro and macro average curves; (e) Shows the results for loss and accuracy whilst using k-fold cross validation, where k = 5. }\label{fig:cnn_output}
\end{figure}

Figure \ref{fig:cnn_output} shows the results from training and testing the model. Figure \ref{fig:cnn_output} (a) and (b) show the results from training, with the training and validation accuracy plotted in (a). It is shown that the network was trained only for 10 epochs to prevent over-fitting. The training accuracy was ~98\% and the validation accuracy was slightly lower at ~94\%, these are excellent accuracies for the number of epochs used. Figure \ref{fig:cnn_output} (b) shows the training and validation loss for the same number of epochs. Both losses fall quite sharply and then start to level off, these could be improved with a larger data set which could be run for more epochs. The loss levelling out indicates that training should be stopped to prevent over-fitting and further improvements can be made from creating larger data sets. To further validate the training process and its outputs, k-fold cross validation was implemented, similar to that implemented by \citet{bobra2015}. The loss and accuracy values from 5-fold cross validation are shown in Figure \ref{fig:cnn_output}(e), with the mean accuracy across the 5 folds being approximately $92.9\% \pm 2.98\%$.

\section{Data preparation}
\label{sec:dataprep}
To create a neural network that can analyse the flare ribbons observed, a robust data set of flaring regions and their ribbons was created. The data set must be created from observations from the same wavelength and instrument to ensure the CNN will not train on varying parameters such as wavelength or smaller features that would perhaps only be found by using a certain instrument. Due to this the data has been collected from the \textit{Atmospheric Imaging Assembly} (AIA) on board the \textit{Solar Dynamics Observatory} (SDO) at the 1600 $\mathring{A}$ wavelength. This wavelength has been chosen as it observes the upper photosphere and transition region allowing for a clearer view of the flare ribbons than those observed in the EUV wavelengths. 

To find dates where flares were observed on the solar disk, the flare ribbon database created by \citet{ribbons} was used. From this database all flares that occurred between November 2012 and December 2014 were included in the training set, this included all C, M and X class flares. To create a training set all of the flares included must be labelled as belonging to a class that is defined for the CNN. Flares where ribbons were not well defined were removed from the data set. This resulted in a training set containing 540 image samples with 160 quiet Sun regions, 160 two-ribbon flares, 95 limb flares and 125 circular flares.

When creating the training and test sets, flares have been chosen such that they should clearly fall into a particular class. To be able to classify each image the following process was implemented. 

For each flare, the observation was chosen at peak flare time according to the \textit{Heliophysics Event Knowledgebase} (HEK) \cite{HEK}. It should be noted that this means the CNN does not take into account the evolution of the flare ribbons from the start to the end of the flare, although this is something that could possibly be included in further work. For some observations there is more than one flare present and in this case both regions are processed and classified separately, although they occurred on the solar disc simultaneously. 

Once the flare position has been located, a bounding box is created around the central flare position. For each flare this creates a bounding box of size $500 \times 500 \times 1$ pixels. This step was included to reduce the size of the data the neural network would have to process due to  large data sets increasing the number of training parameters quickly. The original AIA level 1 data files are $4096 \times 4096 \times 1$ in size, hence this step allows the data input size to be drastically reduced. This code works in a similar way to that of an object detector creating bounding boxes around objects to be classified.

Once located each image is labelled manually according to the classes previously discussed; the quiet sun, two-ribbon flares, circular/compact ribbon flares or limb flares. Once one of these has been chosen, the label is entered into an array ready for training the CNN. 

Once each image has been classified the final steps of the data preparation is to ensure all ROIs were of a suitable size for the CNN to process, hence the data was down-sampled so each image was of size $250 \times 250 \times 1$. Hence the final set of input data would be of size $n \times 250 \times 250 \times 1$, where n is the total number of ROI samples contained within the training data.

The final step for the data preparation was to normalise the data slightly before training, this will ensure the best results when training the CNN and so all of the ROIs were normalised using their z-scores as follows:
\begin{equation*}
    \text{normalised} = \frac{\text{data} - mean(\text{data})}{standard\:deviation(\text{data})}.
\end{equation*}
Once all of the above processes had been carried out on the observations the CNN could begin training as discussed in Section \ref{training}.

\section{Results}
\label{sec:results}
Once training was completed the network was tested using a previously unseen data set. This test set contained 430 images consisting of 160 quiet sun images, 160 two-ribbon flares, 47 limb flares and 63 circular ribbon. Note that some flares included in the test data may have occurred in the same active regions as images included in the training data set. The test outputs are shown using a confusion matrix and ROC curves as shown in Figure \ref{fig:cnn_output} (c) and (d).

A confusion matrix is a good way to visualise model performance on test data that has already been labelled. It summarises the number of correct and incorrect classifications and shows them by plotting the predicted classes against the true classes of the data. The confusion matrix shown in Figure \ref{fig:cnn_output}(c) indicates the percentage of data correctly classified by the diagonal. It shows that for the quiet Sun, two-ribbons and limb classes approximately 95\% of all test data was correctly classified, however for compact flares only 88\% of the data is being correctly classified with approximately 11\% being incorrectly classified as limb flares. This may be due to the distortion of ribbons on the limb, making them look almost compact or circular in shape. The 11\% being incorrectly classified could possibly be corrected by training the model further on a larger data set.

Figure \ref{fig:cnn_output}(d) shows multiple \textit{receiver operating characteristic} (ROC) curves. A ROC curve is plotted as the true positive rate (TPR) against the false positive rate (FPR) at various thresholds. The area under the ROC curve (AUC) indicates the performance of the model as a classifier. The closer to 1 the AUC is indicates how well the model works, with 0 indicating that the model is not classifying anything correctly. hence the further to the left of the diagonal the ROC curve lies the better the classifier. The ROC curves in Figure \ref{fig:cnn_output}(d) show how well the model works for each class, with high AUC values found - all approximately 99\%.

\begin{figure}[ht!]
\begin{center}
\includegraphics[scale=0.45]{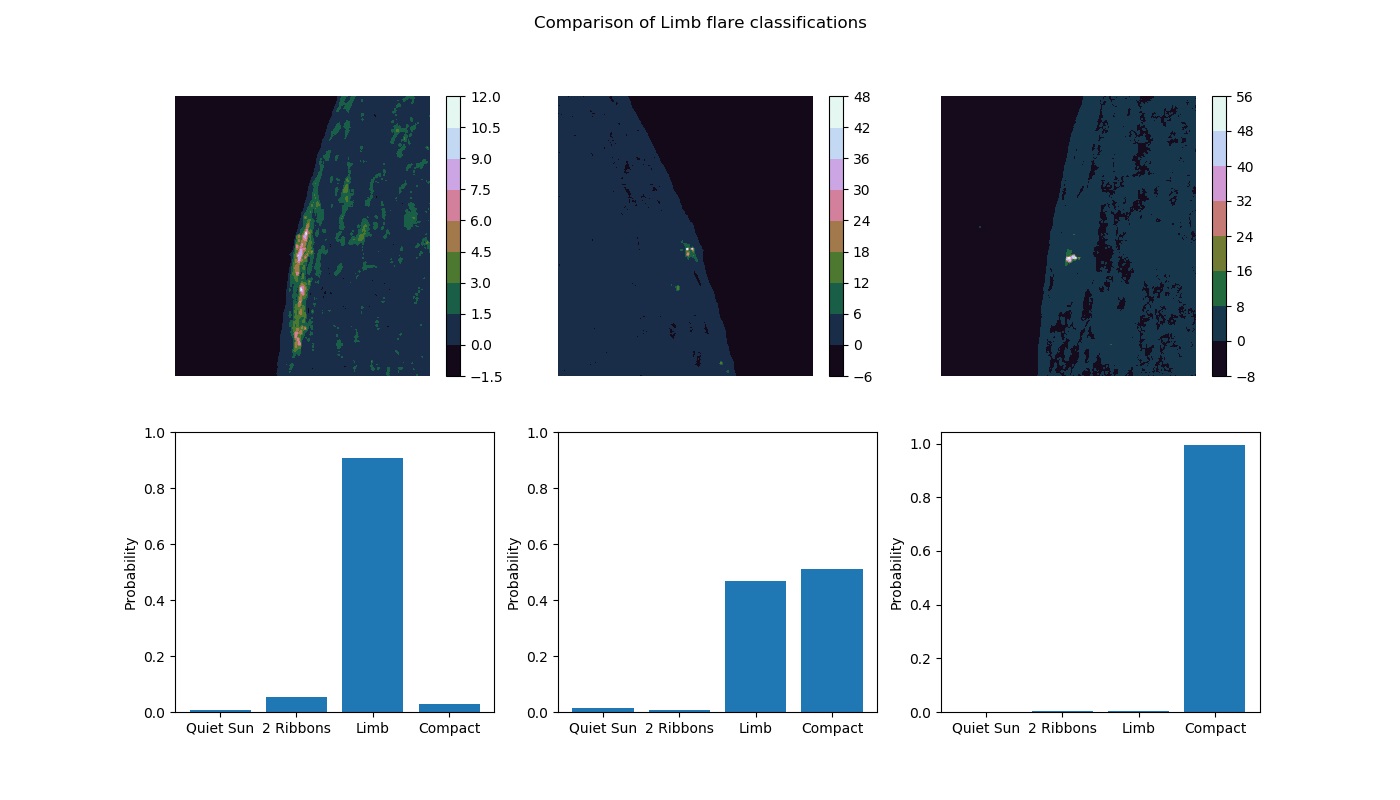}
\end{center}
\caption{ Model output on previously unseen images in the test set. All of the data should belong to the limb flare class, however confusion is seen between limb flares and compact flares.}\label{fig:limb_class}
\end{figure}

To further investigate the model outputs for the limb class, three different images from the test set were considered. Figure \ref{fig:limb_class} shows these three flares and their probabilities of belonging to each class. The first flare is clearly identified as a limb flare with the flaring region sitting just away from the limb. For the second flare it is shown that the model is confused, with very little difference in the confidence that the flare is either a compact or limb flare, both with approximately 50\% probability that the flare could belong to either class. For the final limb flare considered, the model is almost 100\% confident that the flare belongs to the compact/circular ribbon class. This may be due to the flare being slightly further from the limb and so instead of picking up the limb region and the flare, the network has only identified the flare which looks to belong to the circular ribbon class. To rectify this problem in further work some changes to the network and its input could be applied, this could include the inclusion of spatial co-ordinates as one of the inputs which could help with the confusion about which images belong to the limb class.

\section{Discussion}
\label{sec:discussion}
In this paper we have demonstrated a basic application of convolutional neural networks (CNNs) to solar image data. In particular the model classifies the shapes of solar flare ribbons that are visible in 1600$\mathring{A}$ AIA observations. The four classes chosen (Quiet Sun, two-ribbons, Limb flares, Compact/Circular ribbons) were picked due to there being obvious differences between each class, hence more complicated classes could have been chosen but may have effected the overall performance of the CNN. Each of the classes chosen when tested were all found to be well defined with most of the images being correctly classified by the network, with an overall accuracy of approximately 94\%.

The network created is a shallow CNN with only two convolutional layers, unlike deeper networks used on solar image data; \citet{deepcnn_events}, \citet{Armstrong2019}. Both of these papers tried to classify solar events such as flares, coronal holes and sunspots, with varying instruments used. However even with such a shallow CNN as used here, the accuracy of the overall model is still good at approximately 96\%. Our model currently focuses on flare ribbon data and analysing their positions and shapes. This model and data could be compared to a similar setup used to analyse the MNIST dataset containing variations of the numbers 0 to 9, e.g; \citet{mnist_eg}. However to generalise the model further training could be carried out on features such as sunspots or prominences which can also be viewed in the current wavelength, although to do this a deeper network would be needed to extract finer features in the data. Varying the image wavelengths for the AIA data or using a different instrument such as SECCHI EUVI observations from STEREO (\textit{Solar Terrestrial Relations Observatory}) or EIS EUV observations from Hinode could also make the model more robust.

If it was chosen to implement more layers in the network, a CNN such as the VGG network could be used; \citet{VGG}. These networks would take longer to train, particularly on larger data sets containing more images and classes and would require more epochs to properly train the network. As well as increasing the number of convolutional layers used, other layers or parameters could also be modified to alter the model speed and performance. The parameters discussed in Table \ref{table:parameters} could all be altered to affect the model speed and accuracy.

The  main result from this paper shows that even with a shallow convolutional neural network we can get excellent accuracy in the dataset that we considered here. Such a result is encouraging and shows basic CNNs can be very useful tools in analysing large datasets. The model created in this paper can be applied to other data pipelines and can be used to locate many more features from Solar observations obtained from both space and ground-based instruments.

\section*{Conflict of Interest Statement}
The authors declare that the research was conducted in the absence of any commercial or financial relationships that could be construed as a potential conflict of interest.

\section*{Author Contributions}
T.L. created the neural network and carried out the data analysis. T.N. and C.E.P. regularly contributed to the project intellectually by providing ideas and guidance. All authors contributed to the writing of the paper.

\section*{Funding}
T.L acknowledges support by the UK’s Science and Technology Facilities Council (STFC) Doctoral Training Centre Grant ST/P006809/1 (ScotDIST). T.N and C.E.P both acknowledge support by the STFC Consolidated Grant ST/S000402/1.

\section*{Acknowledgments}
The authors would like to thank the continued support from STFC. The AIA data used are provided courtesy of NASA/SDO and the AIA science team.

\section*{Data Availability Statement}
The datasets analysed for this study can be found and downloaded on the JSOC website; http://jsoc.stanford.edu/, with the basic labelled dataset available soon (link will be provided).

\bibliographystyle{frontiersinSCNS_ENG_HUMS} 
\bibliography{test}


\section*{Figure captions}


\begin{table}[ht!]
\centering
 \begin{tabular}{||c|| c c c c||} 
 \hline
 Layer & Number of nodes & Kernel Size(Weights) & Stride & Activation Function\\ [0.5ex] 
 \hline\hline
 Convolution & 32 & $3\times3$ & 1 &ReLU \\ 
 Max Pooling & / & $2\times2$ & 2 & /\\
 Convolution & 64 & $3\times3$ & 1 & ReLU \\
 Max Pooling & / & $2\times2$ & 2 & /\\
 Fully Connected & 128 & $61*61*64 \times 128$ & / &ReLU\\ 
Fully Connected (Output) & 4 & $128 \times 4$ & / & Softmax \\[1.0ex] 
 \hline
\end{tabular}
\caption{Details pf each CNN layer with the number of filters, size of kernels and activation functions used shown.}
\label{table:parameters}
\end{table}

\end{document}